\documentclass[aps,showpacs,showkeys]{revtex4}

\usepackage{amsfonts}
\usepackage{amsmath}
\usepackage{amssymb}
\usepackage{color}
\usepackage{graphicx}
\usepackage[colorlinks=true,allcolors=blue]{hyperref}

\begin{document}

\title{The momentum distribution of two bosons in one dimension with
infinite contact repulsion in harmonic trap gets analytical}
\author{K. Bencheikh\footnote{ORCID: 0000-0002-2867-0706}}
\email{bencheikhkml@univ-setif.dz}
\affiliation{D\'{e}partement de Physique. Laboratoire de physique 
quantique et syst\`{e}mes dynamiques. Universit\'{e} Ferhat Abbas S\'{e}tif-1, 
Setif 19000, Algeria}
\author{L. M. Nieto\footnote{ORCID: 0000-0002-2849-2647}}
\email{luismiguel.nieto.calzada@uva.es}
\affiliation{Departamento de F\'{\i}sica Te\'{o}rica, At\'{o}mica y \'{O}ptica and IMUVA,
Universidad de Valladolid, 47011 Valladolid, Spain}
\author{L. U. Ancarani\footnote{ORCID: 0000-0002-0503-3288}}
\email{ugo.ancarani@univ-lorraine.fr}
\affiliation{Universit\'{e} de Lorraine-CNRS, UMR 7019, LPCT, Metz, 57000, France}
\date{\today}

\begin{abstract}
For a harmonically trapped system consisting of two bosons in one spatial
dimension with infinite contact repulsion (hard core bosons), we derive an
expression for the one-body density matrix $\rho_B$ in terms of centre of
mass and relative coordinates of the particles. 
The deviation from $\rho_F$, the density matrix for the two
fermions case, can be clearly identified. Moreover, the obtained 
$\rho_B$ allows us to derive a
closed form expression of the corresponding momentum distribution $n_{B}(p)$. 
We show how the result deviates from the noninteracting fermionic case,
the deviation being associated to the short range character of the
interaction. Mathematically, our analytical momentum distribution is
expressed in terms of one and two variables confluent hypergeometric
functions. Our formula satisfies the correct normalization and possesses the
expected behavior at zero momentum. It also exhibits the high momentum 
$1/p^4 $ tail with the appropriate Tan's coefficient. Numerical results
support our findings.
\end{abstract}

\maketitle


\section{Introduction}

Two-body models have the merit that they can be solved exactly and give
direct access to the wave function and to the one-body density matrix. We
consider here two bosonic atoms with mass $m$ in one dimension with infinite
repulsive contact interaction (two hard core bosons or the so-called
Tonks-Girardeau regime). The system is also subjected to an external
harmonic potential trap with frequency $\omega$. A system of Tonks-Girardeau
(TG) gas has been already realized in ultra-cold experiments \cite{Kinoshita2004}. 
On the theoretical side the TG model is exactly solvable
through the Bose-Fermi mapping theorem, which relates this gas to a system
of noninteracting spin polarized fermions \cite{Girardeau1960,Girardeau2000}. 
The two systems - bosonic and fermionic - exhibit some identical
properties such as the particle density. However, their corresponding
one-body density matrices are not the same resulting into a considerable
difference between the momentum density of TG gas from that of an ideal
Fermi gas \cite{Wright}. 
While it is rather easy to calculate the momentum density of an ideal 
system of fermions, the task is difficult for a TG gas.
The main
purpose of the present work is to provide a definitive solution for the case
of two particles, deriving a closed form expression of the momentum
distribution for two harmonically trapped interacting bosons in the TG
regime. The analytical approach allows us to clearly identify the deviation
from the fermionic counterpart.

Let us denote by $x_{1},x_{2}$ the positions of the two interacting atoms in
one dimension. According to the Bose-Fermi mapping theorem 
\cite{Girardeau2000}, the ground state wave function of the system $\psi
_{B}(x_{1},x_{2})$ can be written as 
\begin{equation}
\psi _{B}(x_{1},x_{2})=|\psi _{F}(x_{1},x_{2})|  \label{00}
\end{equation}
where $\psi _{F}(x_{1},x_{2})$ is the fermionic ground state, that is, a
Slater determinant 
\begin{equation}
\psi _{F}(x_{1},x_{2})=\frac{1}{\sqrt{2}}\left( \varphi _{0}(x_{1})\varphi
_{1}(x_{2})-\varphi _{0}(x_{2})\varphi _{1}(x_{1})\right) .  \label{1}
\end{equation}
built on the two lowest single particle normalized eigenfunctions $\varphi
_{0}(x)$ and $\varphi _{1}(x)$ of the harmonic oscillator potential trap.
Explicitly we have 
\begin{equation}
\varphi _{0}(x)=\sqrt{\frac{1}{\ell \sqrt{\pi }}}\ e^{-\frac{1}{2}({x}/{\ell 
})^{2}},\qquad \varphi _{1}(x)=\sqrt{\frac{2}{\ell \sqrt{\pi }}}\ ({x}/{\ell 
})\,e^{-\frac{1}{2}({x}/{\ell })^{2}},  \label{1coma5}
\end{equation} 
where $\ell =\sqrt{\hbar /m\omega }$ denotes the harmonic oscillator
characteristic length. After insertion in (\ref{00}), we get 
\begin{equation}
\psi _{B}(x_{1},x_{2})=\frac{1}{\ell \sqrt{\pi }}\ e^{-\frac{1}{2}\left( ({ 
x_{1}}/{\ell })^{2}+({x_{2}}/{\ell })^{2}\right) }\left \vert \frac{x_{1}}{ 
\ell }-\frac{x_{2}}{\ell }\right \vert =\psi _{B}(x_{2},x_{1}).  \label{2}
\end{equation} 
A few words about this wave function are in order. It may also be considered
in the context of one dimensional spinor quantum gases studies, a topic
which has recently received a lot of attention \cite{Deuretzbacher2014,Volosniev2014,Volosniev2015}.
Let us consider the particular case of two ultra-cold bosons in the Tonks-Girardeau regime
occupying or
prepared in two hyperfine states; this constitutes a pseudo-spin-1/2 system
of different spins (which we denote by $\uparrow $ and $\downarrow $), the
wave function $\psi _{B}(x_{1},x_{2})=|\psi _{F}(x_{1},x_{2})|$ in (\ref{2})
being the spatial wave function of a spin-$1/2$ hardcore bosons corresponding
to the full wave function of the system, $\Phi =|\psi
_{F}(x_{1},x_{2})|\otimes \chi ^{S=1}$, where $\chi ^{S=1}=\left( \left \vert
\uparrow \downarrow \right \rangle +\left \vert \downarrow \uparrow
\right \rangle \right) /\sqrt{2}$ is the triplet (symmetric) spin state 
\cite{Alam2020}. Another possible configuration \cite{Alam2020} would be to use
a singlet (antisymmetric) spin state $\chi ^{S=0}=\left( \left \vert \uparrow
\downarrow \right \rangle -\left \vert \downarrow \uparrow \right \rangle
\right) /\sqrt{2}$. In this case the full wave function would simply read 
$\Psi =\psi _{F}(x_{1},x_{2})\otimes \chi ^{S=0}$, and calculations are very
easy. In the present investigation we consider the much more difficult
triplet spin state case for which the calculations are by far not
straightforward because of the presence of the absolute value in the wave
function.

The plan of the work is the following. In sect.~\ref{section2} we use the two-particle
wave function in (\ref{2}) to derive an expression of the one-body density
matrix in terms of centre of mass and relative coordinates. This first
result is then used in sect.~\ref{section3} to derive a closed form for the momentum
density which holds for arbitrary values of the momentum $p$. This second
result is analyzed and tested in sect.~\ref{section4} along three lines: (i) we check
that one recovers the correct value of the momentum density at $p=0$; (ii)
we check that our density satisfies the required normalization condition 
$\int n_{B}(p)dp=2$; (iii) we prove analytically, and we verify numerically,
that our momentum distribution exhibits a well-known result in quantum gases
with contact repulsive interactions \cite{Olshanii2003}, namely that the
momentum distribution of the atoms decays asymptotically as $1/p^{4}$ for
large momentum $p$, and as a consequence we derive the exact expression of
the so-called Tan's contact coefficient for the case of two interacting
particles \cite{Tan}. In the last section the results of the paper are
summarized.

\section{One-body density matrix of a one-dimensional system of two bosons
with infinite contact repulsion in harmonic trap}\label{section2}

The one body density matrix for the system under study is given by 
\cite{Dreizler1990} 
\begin{equation}
\rho _{B}(x_{1},x_{2})=2\int_{-\infty }^{+\infty }\psi _{B}(x_{1},x^{\prime
})\psi _{B}^{\ast }(x_{2},x^{\prime })\ dx^{\prime } .  \label{3}
\end{equation}
Using the explicit expression (\ref{2})  and denoting $z_1= x_1/\ell$, 
$z_2=x_2/\ell$, $\xi=x^{\prime }/\ell$, the density can be written as 
\begin{eqnarray}
\rho _{B}(x_{1},x_{2})&=&\frac{2}{\ell \pi } \ e^{-\frac12 (z_1^2+z_2^2)}
\int_{-\infty }^\infty e^{-\xi^2} \ |z_1-\xi |\, |z_2-\xi| \, d\xi =\frac{2}{
\ell \pi}\ e^{-\frac12 ( z_1^2+z_2^2 )} \int_{-\infty }^{\infty }
e^{-\xi^2}\, \left| (\xi- u )^{2}-v^2 \right| d\xi ,  \label{5}
\end{eqnarray}
where $u$ is the dimensionless coordinate of the centre of mass of the two
particles and $v$ is half the dimensionless relative coordinate of the two
particles: 
\begin{equation}
u=\frac{z_{1}+z_{2}}{2}=\frac{x_{1}+x_{2}}{2\ell}, \qquad v=\frac{z_{1}-z_{2} 
}{2}=\frac{x_{1}-x_{2}}{2\ell}.  \label{uv}
\end{equation}
Performing now the change of variable $w=(\xi-u)$, eq.~(\ref{5}) takes
the form 
\begin{equation}
\rho _{B}(x_{1},x_{2})=\frac{2}{\ell \pi }\, e^{-(u^{2}+v^{2})}\int_{-\infty
}^\infty e^{- (w+u)^2} \left| w^2-v^2\right| \, dw.  \label{6}
\end{equation}
To carry out the above integration, we remove the absolute value symbol and
write the density matrix as 
\begin{equation}
\rho _{B}(x_{1},x_{2})=\frac{2}{\ell \pi }\, e^{-(u^{2}+v^{2})}\, \mathcal{A}
(u,v), \qquad \mathcal{A} (u,v)= \mathcal{B}(u,v)- \mathcal{C} (u,v) + 
\mathcal{B} (-u,v),  \label{7}
\end{equation}
where 
\begin{equation}
\mathcal{B}(u,v)=\int_{-\infty }^{-|v|} e^{- (w+u)^2}\, (w^2-v^2)\, dw,
\qquad \mathcal{C} (u,v)=\int_{-|v|}^{|v|} e^{- (w+u)^2} \, (w^2-v^2) \, dw .
\label{11}
\end{equation}
The three functions $\mathcal{A}(u,v)$, $\mathcal{B}(u,v)$ and $\mathcal{C}
(u,v)$ are calculated in Appendix~\ref{appendixa} in terms of the error function 
$\text{erf}(x)=2/\sqrt{\pi }\int_{0}^{x}e^{-t^{2}}dt$. Substituting the result 
(\ref{A11}) into (\ref{7}), we obtain the one-body density matrix $\rho
_{B}(x_{1},x_{2})$ in the form 
\begin{eqnarray}
\rho_B(x_1,x_2)\!&\!=\!&\! \frac{2}{\ell \pi}e^{-\left( u^2+{v^2}\right)}  
\Big \{ \left( u+{|v|}\right) e^{-\left( u-{|v|}\right)^2}-\left( u-{|v|} 
\right) e^{-\left(u+{|v|}\right)^2}  \label{14} \\
&&\qquad \qquad \qquad +\sqrt{\pi} \left( \tfrac{1}{2}-v^2+u^2\right) \left[
1+\text{erf}\left( u-{|v|}\right)-\text{erf}\left(u+{|v|}\right) \right]  
\Big \} .  \notag
\end{eqnarray}

For two noninteracting fermions, the one-body density matrix is given by 
\begin{equation}
\rho _{F}(x_{1},x_{2})=\varphi _{0}(x_{1})\varphi _{0}^{\ast
}(x_{2})+\varphi _{1}(x_{1})\varphi _{1}^{\ast }(x_{2})=\frac{2}{\ell 
\sqrt{\pi}}\, e^{-\left( u^2+{v^2}\right)} \left( \tfrac{1}{2}-v^2+u^2\right) ,
\label{15}
\end{equation} 
where we have used the explicit expressions (\ref{1coma5}) of $\varphi
_{0}(x)$, $\varphi _{1}(x)$. Remark that these densities are correctly
normalized to the particle number, so that $\int \rho_B(x,x)dx=\int
\rho_{F}(x,x)dx=2$. By comparing the bosonic to the fermionic density, we
may write 
\begin{eqnarray}
\rho _{B}(x_{1},x_{2}) = \rho _{F} (x_{1},x_{2})+ \rho _{D} (x_{1},x_{2})
\label{BFD}
\end{eqnarray}
and thus clearly identify the deviation 
\begin{eqnarray}
\rho _{D}(x_{1},x_{2}) &=& \frac{2}{\ell \pi} \left[ 2u \sinh \left( 2u |v|
\right) +2|v| \cosh \left( 2u |v|\right) \right] \, e^{-2u^2}e^{-2v^2} 
\notag \\
&&+ \frac{2}{\ell \sqrt{\pi}} \left( \tfrac{1}{2}-v^2+u^{2} \right) \left[ 
\text{erf}\left( u-|v|\right) -\text{erf}\left( u+|v|\right) \right]\,
e^{-u^2} e^{-v^2} .  \label{deviation}
\end{eqnarray}
Note that if $x_1=x_2=x$ then $u=x$, $v=0$, and we obtain $\rho_D(x,x)=0$
and therefore $\rho _{B}(x,x)=\rho _{F}(x,x)$, as required by the Bose-Fermi
mapping theorem \cite{Girardeau2000}.

Expression (\ref{BFD}), together with the explicit deviation  
(\ref{deviation}), constitute the first outstanding result of the present
study. It should be pointed out that, although the Tonks-Girardeau gas and
the ideal Fermi gas have identical local spatial density $\rho _{B}(x)=\rho
_{F}(x)$, they display different one-body density matrices yielding to very
different momentum density profiles \cite{Pezer2007}. In the sequel, this
property will be explicitly shown and analytically verified in the case of
two particles.

Before proceeding further, it is interesting to note that, at this level,
one can perform an expansion of the density matrix in (13) with (14) in
terms of $|v|=|x_{1}-x_{2}|/(2\ell )$. This corresponds to a short-distance
expansion and allows one to directly access to momentum density profile at
momentum high-values \cite{Boumaza-Doct}. This asymptotic behaviour is of
particular interest in the study of short-range interacting systems \cite{Vignolo2013}.

\section{Closed-form expression of the momentum distribution}\label{section3}

In what follows we will derive an analytical expression of the momentum
distribution for the system of two bosons in the Tonks-Girardeau regime. 
In terms of the one-body density matrix, the momentum density denoted by 
$n_{B}\left( p\right) $ is given by \cite{Dreizler1990} 
\begin{equation}
n_{B} ( p ) =\frac{1}{2\pi \hbar }\int_{-\infty}^\infty
\int_{-\infty}^\infty \rho _{B} (x_{1},x_{2} ) \, e^{-ip ( x_1-x_2)/\hbar}
\, dx_{1}dx_{2}.  \label{20}
\end{equation} 
and is normalized to the total particle number, i.e., $\int n_{B}\left(
p\right) dp=2$. 
In terms of centre of mass and relative dimensionless coordinates $(u,v)$,
we have 
\begin{equation}
n_{B}\left( p\right) =\frac{\ell^2}{\pi \hbar } \int_{-\infty}^\infty
\int_{-\infty}^\infty \rho _{B}\left( \ell(u+v),\ell(u-v)\right) \,
e^{-2i\ell pv/\hbar}\, du\,dv .  \label{21}
\end{equation} 
Now, substituting (\ref{BFD}) into (\ref{21}), we can write 
\begin{equation}
n_{B}\left( p\right) =n_{F}(p)+n_{D}(p)  \label{25}
\end{equation} 
with 
\begin{equation}
n_{F}(p)=\frac{2\ell}{\sqrt{\pi}\pi \hbar} \int_{-\infty}^\infty
\int_{-\infty}^\infty \left( \tfrac{1}{2}-v^2+u^2\right)
e^{-u^2}e^{-v^2}e^{-2i\ell pv/\hbar} du\, dv,  \label{26}
\end{equation} 
which is nothing but the momentum density of two noninteracting fermions in
harmonic trap, and 
\begin{eqnarray}
n_{D}(p) &=& I_1+I_2= \frac{4\ell}{\pi^2 \hbar}
\int_{-\infty}^\infty \int_{-\infty}^\infty \bigl( \left \vert v\right \vert
\cosh \left( 2u|v|\right) + u \sinh \left( 2u|v| \right) \bigr)  
e^{-2u^2}e^{-2v^2} e^{-2i\ell pv/\hbar}\, du\, dv  \notag \\
&&+\frac{2\ell }{\pi^{3/2} \hbar} \int_{-\infty}^\infty
\int_{-\infty}^\infty \left( \tfrac{1}{2}-v^2+u^2 \right) \bigl( \text{erf} 
\left( u -|v| \right) -\text{erf}\left( u+|v| \right) \bigr) e^{-u^2}\,
e^{-v^2} e^{-2i\ell pv/\hbar} \, du\, dv ,  \label{28}
\end{eqnarray} 
which is the momentum density corresponding to the deviation $\rho_D$. Next
we are going to evaluate the integrals in (\ref{26})--(\ref{28}). It will be
convenient to use the momentum dependent dimensionless variable $q=\ell
p/\hbar$.

\subsection{Evaluation of $n_{F}(p)$}

The integration with respect to $u$ in (\ref{26}) yields 
\begin{equation}
n_{F}(p)=\frac{4\ell }{\hbar \pi }\left[ \int_{0}^{\infty }e^{-v^{2}}\cos
(2qv)\,dv-\int_{0}^{\infty }v^{2}e^{-v^{2}}\cos (2qv)\,dv\right]   \label{30}
\end{equation} 
By using the eq.~(3.952), number
(9) of ref.~\cite{Gradshteyn--Ryzhik}
\begin{equation}
\int_{0}^{\infty }x^{2n}\,e^{-\beta ^{2}x^{2}}\, \cos (\alpha x)\,dx=
\frac{ (-1)^{n}\sqrt{\pi }}{(2\beta )^{2n+1}}\, e^{-\frac{\alpha ^{2}}{4\beta ^{2}}}
\, H_{2n}\left( \frac{\alpha }{2\beta }\right) ,  \label{31}
\end{equation} 
the integrals in (\ref{30}) are easily computed, giving the final expression
of the momentum density $n_{F}(p)$: 
\begin{equation}
n_{F}(p)=\frac{\ell }{\sqrt{\pi }\hbar }\,(1+2q^{2})\,e^{-q^{2}}.  \label{32}
\end{equation} 
It is easy to check that it is normalized as $\int n_{F}\left( p\right) dp=2$. 
The same $n_{F}(p)$ result can be derived also 
by using the expression, $n_{F}(p)=\left \vert \widetilde{\varphi } 
_{0}(p)\right \vert ^{2}+\left \vert \widetilde{\varphi }_{1}(p)\right \vert
^{2}$, where $\widetilde{\varphi }_{0,1}(p)$ denote the single particle wave
functions in momentum representation, i.e., the Fourier transforms 
\begin{equation*}
\widetilde{\varphi }_{0,1}(p)=\frac{1}{\sqrt{2\pi \hslash }}\int \varphi
_{0,1}(x)e^{-i\frac{p}{\hbar }x}dx
\end{equation*} 
of the single particle wave functions $\varphi _{0,1}(x)$ \cite{Tannoudji1977}.

\subsection{Evaluation of $n_{D}(p)$}

Let us start with the second term of the first double integral 
$I_1$ in (\ref{28}), and perform an integration by parts with respect to $u$,
to get 
\begin{equation}
\int_{-\infty }^{\infty }\int_{-\infty }^{\infty } u\sinh \left( 2u\left
\vert v\right \vert \right) e^{-2u^{2}}e^{-2 v^{2}} e^{-2iqv}\, du\, dv= 
\frac{1}{2}\int_{-\infty }^{\infty }\int_{-\infty }^{\infty } \left \vert
v\right \vert \cosh \left( 2u\left \vert v\right \vert \right)
e^{-2u^{2}}e^{-2v^2} e^{-2iqv}\, du\, dv .  \label{34}
\end{equation} 
Combining now the two terms, and integrating over $u$ (simple Gaussian-type
integral \cite{Gradshteyn--Ryzhik}), the first double integral in (\ref{28})
simplifies into 
\begin{equation}
I_1=\frac{6\ell}{\pi ^{2}\hbar }\int_{-\infty }^{\infty
}\int_{-\infty }^{\infty } \left \vert v\right \vert \cosh \left( 2u\left
\vert v\right \vert \right) e^{-2u^{2}}e^{-2v^2}e^{-2iqv}\, du\, dv 
=\frac{6\ell \sqrt{2}}{\pi ^{3/2}\hbar }\int_{0}^{\infty }ve^{-\frac{3}{2}v^{2}}
\cos \left( 2qv\right) \, dv .  \label{36}
\end{equation} 
For the second double integral $I_2$ in (\ref{28}), we use the
property $\text{erf}(-x)=-\text{erf}(x)$, to write 
\begin{equation}
I_2= \frac{4\ell}{\pi ^{3/2}\hbar }\int_{-\infty }^{\infty
}e^{-v^2} \cos(2qv)\, dv \int_{-\infty }^{\infty }\left( \tfrac{1}{2}-v^2
+u^{2}\right) \, \text{erf}\left( u-|v|\right) \, e^{-u^{2}}du .  \label{38}
\end{equation}
Although it is not at all straightforward because of the presence of the
error function, we were able to proceed again by reducing the double
integral (\ref{38}) to a single one (see result (\ref{B14}) in Appendix~\ref{appendixb}): 
\begin{eqnarray}
 I_2& =&\frac{4\ell}{\pi ^{3/2}\hbar }\left \{ -2%
\sqrt{\pi }\int_{0}^{\infty }e^{-v^{2}}\cos (2qv)\, \text{erf}\left( \frac{v%
}{\sqrt{2}}\right) dv +2\sqrt{\pi} \int_{0}^{\infty }v^{2}e^{-v^2}\cos (2qv)
\, \text{erf}\left( \frac{v}{\sqrt{2}}\right) dv \right.  \notag \\
&&\qquad \qquad + \left. \frac{\sqrt{2}}{2}\int_{0}^{\infty
}e^{-3v^2/2}\,v\, \cos (2qv)\, dv \right \}.  \label{39}
\end{eqnarray}
Collecting the results (\ref{36}) and (\ref{39}), the momentum density %
(\ref{28}) may be written as 
\begin{eqnarray}
n_{D}(p) =\frac{8\ell}{\pi ^{3/2}\hbar }\left \{ \sqrt{2}\, J_A -\sqrt{\pi }
\, J_B+ \sqrt{\pi }\, J_C \right \} ,  \label{40}
\end{eqnarray}
in terms of three single integrals 
\begin{eqnarray}
J_{A} &=&\int_{0}^{\infty }e^{-3v^2/2}v\cos \left( 2qv\right) dv ,
\label{41} \\
J_{B} &=&\int_{0}^{\infty }e^{-v^2}\cos \left( 2qv\right) \text{erf}\left( 
\frac{v}{\sqrt{2}}\right) dv ,  \label{42} \\
J_{C} &=&\int_{0}^{\infty }e^{-v^2}v^{2}\cos \left( 2qv\right) \text{erf}%
\left( \frac{v}{\sqrt{2}}\right) dv .  \label{43}
\end{eqnarray}%
we proceed with their evaluation.

For $J_{A}$, we first perform an integration by parts to write 
\begin{equation}
J_{A}=\frac{1}{3}\left[1-2q\int_{0}^{\infty }e^{-3v^2/2} \sin \left(
2qv\right) dv\right] .  \label{45}
\end{equation}%
We then apply the identity (see eq. (3.896), number (3) of ref.~\cite{Gradshteyn--Ryzhik}) 
\begin{equation}
\int_{0}^{\infty }e^{-\alpha x^{2}}\sin \left( \beta x\right) dx=\frac{\beta 
}{2\alpha }M\left( 1,\frac{3}{2};-\frac{\beta ^{2}}{4\alpha }\right) ,
\label{46}
\end{equation}%
where $M\left( a,b;z\right) \equiv $ $_{1}F_{1}\left( a,b;z\right) $ is the
confluent hypergeometric function defined by \cite{Gradshteyn--Ryzhik} 
\begin{equation}
M\left( a,b;z\right) =\sum \limits_{n=0}^{\infty }\frac{(a)_{n}}{(b)_{n}}%
\frac{z^{n}}{n!} ,  \label{48}
\end{equation}%
being $(a)_{n}=a(a+1)(a+2)....(a+n-1)=\Gamma(a+n)/\Gamma(a)$ the Pochhammer
symbol. We thus get 
\begin{equation}
J_{A}=\frac{1}{3}\left[ 1-\frac{4}{3}\, q^{2}\, M\left( 1,\frac{3}{2};-\frac{%
2}{3} \, q^{2} \right) \right] .  \label{47}
\end{equation}

For $J_{B}$, we first replace $\cos(x)=(e^{ix}+e^{-ix})/2$, and then use the
tabulated relation \cite{Prudnivkov2002}%
\begin{eqnarray}
\int_{0}^{\infty }x^{\alpha -1}e^{-\beta x}e^{-\gamma x^{2}}\text{ }%
(\varepsilon x)dx &=&\frac{\varepsilon }{\sqrt{\pi }\gamma ^{\frac{\alpha +1%
}{2}}}\Gamma \left( \frac{\alpha +1}{2}\right) \Psi _{1}\left( \frac{\alpha
+1}{2},\frac{1}{2},\frac{3}{2},\frac{1}{2};-\frac{\varepsilon ^{2}}{\gamma },%
\frac{\beta ^{2}}{4\gamma }\right)  \notag \\
&&- \frac{\varepsilon \beta }{\sqrt{\pi }\gamma ^{\frac{\alpha }{2}+1}}%
\Gamma \left( \frac{\alpha }{2}+1\right) \Psi _{1}\left( \frac{\alpha }{2}+1,%
\frac{1}{2},\frac{3}{2},\frac{3}{2};-\frac{\varepsilon ^{2}}{\gamma },\frac{%
\beta ^{2}}{4\gamma }\right) ,  \label{49}
\end{eqnarray}%
where $\Psi _{1}\left( a,b,c_{1},c_{2};z_{1},z_{2}\right) $ is the
two-variable (degenerate) confluent hypergeometric series \cite%
{Prudnivkov2002,Strivastava1985}, given by \cite{Ancarani2017}: 
\begin{eqnarray}
\Psi _{1}\left( a,b,c_{1},c_{2};z_{1},z_{2}\right) &=&\sum
\limits_{m=0}^{\infty }\sum \limits_{n=0}^{\infty }\frac{(a)_{m+n}(b)_{m}}{%
(c_{1})_{m}(c_{2})_{n}}\frac{z_{1}^{m}}{m!}\frac{z_{2}^{n}}{n!} , =\sum
\limits_{m=0}^{\infty }\frac{(a)_{m}(b)_{m}}{(c_{1})_{m}}\frac{z_{1}^{m}}{m!}%
M\left( a+m,c_{2};z_{2}\right)  \label{50} \\
&=& \sum \limits_{n=0}^{\infty }\frac{(a)_{n}}{(c_{2})_{n}}\frac{z_{2}^{n}}{%
n!}\text{ }_{2}F_{1}\left( a+n,b,c_{1;}z_{1}\right) , \qquad \left \vert
z_{1}\right \vert <1,  \label{52}
\end{eqnarray}%
where $_{2}F_{1}(a,b,c;z)=\sum \nolimits_{n=0}^{\infty }\left({(a)_{n}(b)_{n}%
}/{(c)_{n}}\right) {z^{n}}/{n!}$ is the Gaussian hypergeometric function. 
In our case $\alpha =1$, $\beta =\mp iq,\gamma =1/4$, $\varepsilon =$ $1/(2%
\sqrt{2})$, and replacing $q=\ell p/\hbar$, we have 
\begin{equation}
J_{B}=\frac{1}{2}\int_{0}^{\infty }e^{-\frac{v^{2}}{4}}e^{iqv}\, \text{erf}%
\left( \frac{v}{2\sqrt{2}}\right) dv+\frac{1}{2}\int_{0}^{\infty }e^{-\frac{%
v^{2}}{4}}e^{-iqv} \, \text{erf}\left( \frac{v}{2\sqrt{2}}\right) dv = \frac{%
1}{2} \sqrt{\frac{2}{\pi }}\Psi _{1}\left( 1,\frac{1}{2},\frac{3}{2},\frac{1%
}{2};-\frac{1}{2},\frac{-\ell^{2}p^{2}}{\hbar ^{2}}\right) .  \label{54}
\end{equation}%
The evaluation of $J_{C}$ is similar with $\alpha =3$ instead of $\alpha =1$ 
\begin{equation}
J_{C}=\frac{1}{2} \sqrt{\frac{2}{\pi }}\Psi _{1}\left( 2,\frac{1}{2},\frac{3%
}{2},\frac{1}{2};-\frac{1}{2},-\frac{\ell^{2}p^{2}}{\hbar ^{2}}\right) .
\label{55}
\end{equation}%
Note that $J_{B}$ and $J_{C}$ converge as functions of the momentum $p$, as
in both cases the function $\Psi _{1}$ has $z_{1}=-1/2$, which is within the
convergence interval $\left \vert z_{1}\right \vert <1$.

Hence, collecting the results, we find 
\begin{eqnarray}
n_{D}(p) =\frac{4\ell}{\pi \hbar }\sqrt{\frac{2}{\pi }}\left \{ \Psi
_{1}\left( 2,\frac{1}{2},\frac{3}{2},\frac{1}{2};-\frac{1}{2},-\frac{%
\ell^{2}p^{2}}{\hbar ^{2}}\right) -\Psi _{1}\left( 1,\frac{1}{2},\frac{3}{2},%
\frac{1}{2};-\frac{1}{2},-\frac{\ell^{2}p^{2}}{\hbar ^{2}}\right) \right.
    \nonumber \\
+\left. \frac{2\ }{3}\left[ 1-\frac{4}{3}\left( \frac{\ell p}{\hbar }%
\right)^{2} M\left( 1,\frac{3}{2};-\frac{2}{3}\frac{\ell^{2}p^{2}}{\hbar ^{2}%
}\right) \right] \right \} .  \label{56}
\end{eqnarray}

\subsection{Closed-form expression of the momentum distribution $n_{B}\left(
p\right) $}

Collecting $n_{F}(p)$ from (\ref{32}) and the result in (\ref{56}), we
obtain the closed-form expression of the momentum distribution 
\begin{eqnarray}  
n_{B}\left( p\right) &=&\frac{\ell}{\sqrt{\pi }\hbar }\left[ 1+ \frac{%
2\ell^{2}p^{2}}{\hbar ^{2}}\right] e^{\frac{-\ell^{2}p^{2}}{\hbar ^{2}}} +%
\frac{\ell}{\sqrt{\pi }\hbar } \left( \frac{4\sqrt{2}}{\pi }\right) \left \{ 
\frac{2\ }{3}\left[ 1-\frac{4}{3}\left( \frac{\ell p}{\hbar }\right)
^{2}M\left( 1,\frac{3}{2};-\frac{2}{3}\frac{\ell^{2}p^{2}}{\hbar ^{2}}%
\right) \right] \right.  \nonumber  \\
&& \qquad\qquad\qquad \left. +\Psi _{1}\left( 2,\frac{1}{2},\frac{3}{2},\frac{1}{2};-\frac{1}{2},-%
\frac{\ell^{2}p^{2}}{\hbar ^{2}}\right) -\Psi _{1}\left( 1,\frac{1}{2},\frac{%
3}{2},\frac{1}{2};-\frac{1}{2},-\frac{\ell^{2}p^{2}}{\hbar ^{2}}\right)
\right \}.  \label{57}
\end{eqnarray}
which is the principal result of our paper. In the next section, we provide
strong tests that support its validity.

\begin{figure}[htb]
\centering
\includegraphics[width=0.45\textwidth]{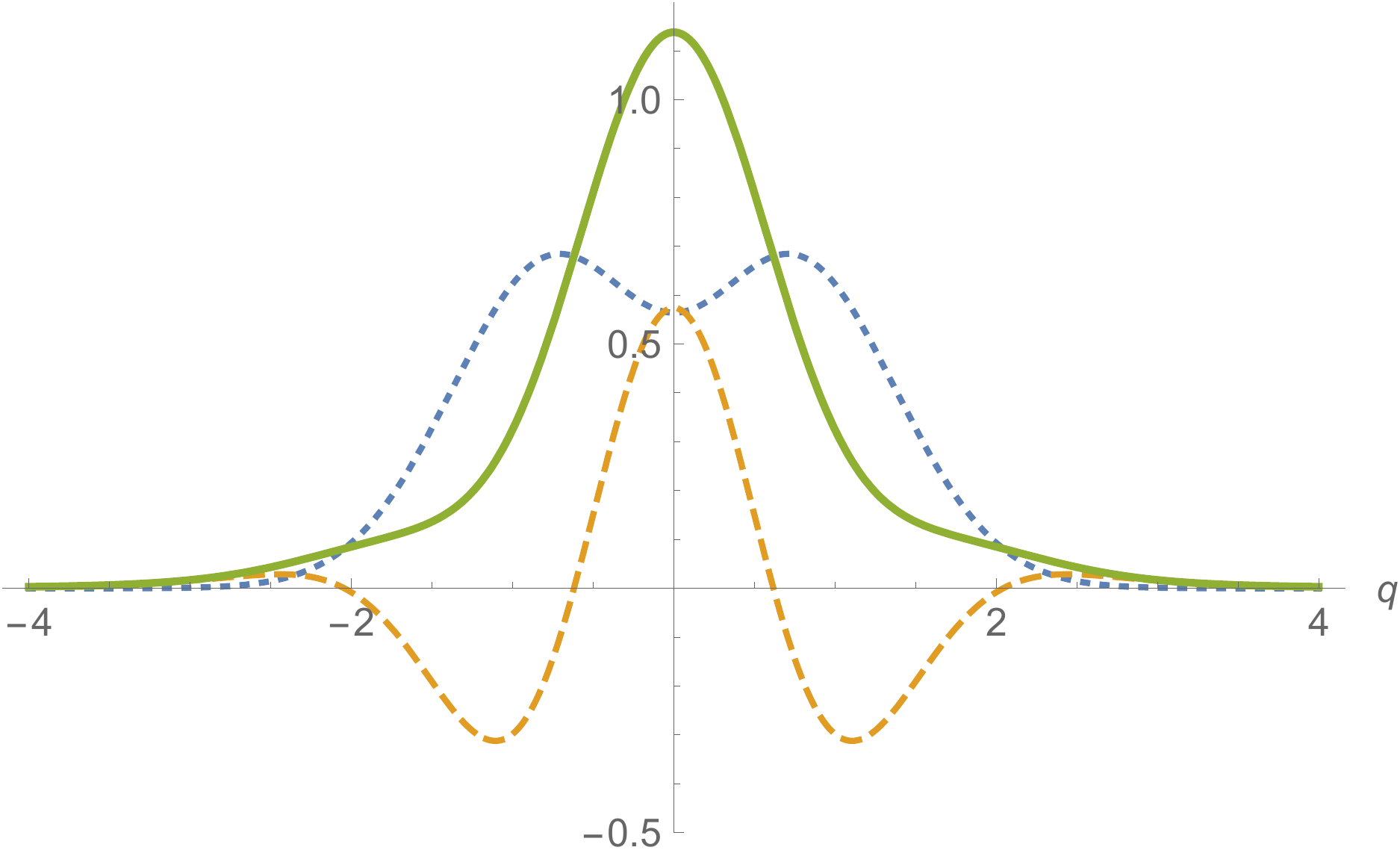}
\caption{Plot of $n_{B}(p) \hbar/\ell$ as a function of $q=\ell
p/\hbar$ (green solid curve) as given by (\ref{57}). We also display $%
n_{F}(p) \hbar/\ell$ in (\ref{32}) (blue dotted curve) and their difference $%
n_{D}(p)\hbar/\ell$ (\ref{56}) (orange dashed curve).}
\label{fig1}
\end{figure}

We have computed $n_B(p)$ for any value of $p$ using either of the two
series representations (\ref{50}) or (\ref{52}). For the evaluation of the $%
\Psi_1$ function, from a numerical convergence point of view it turns out
that the series in terms of confluent hypergeometric functions is more
convenient because $\vert z_1 \vert <1$ while $\vert z_2 \vert$ can take any
positive value. The numerical results for $n_B(p)$ has been further
confirmed by using (\ref{40}) and direct numerical integration of (\ref{41}), 
(\ref{42}) and (\ref{43}). A plot of $\hbar n_{B}/\ell$ as a function of $%
q={\ell p}/{\hbar }$ is shown in fig.~\ref{fig1}, and is compared to both
contributions $n_{F}$ and $n_{D}$. We clearly observe a substantial shape
difference between the fermionic and bosonic cases. The fermionic
distribution presents a minimum at $q=0$, a maximum at $q=1/\sqrt{2}$, and
decreases rapidly to zero due to the $e^{-q^2}$ term. The bosonic
distribution, on the other hand, has almost double value at $q=0$ (see ratio 
$n_{B}\left( 0\right) /n_{F}(0)\approx 2.02$ discussed in sect.~\ref{section4a}), and
decreases thereafter but in much slower fashion; approximately beyond $q>3$,
we have $n_F\simeq 0$ and $n_B$ is essentially due to the deviation $n_D$.

\section{Tests of the momentum distribution expression (\ref{57})}\label{section4}

We consider now three tests on the analytical expression of the momentum
distribution $n_B(p)$. First, we will verify that the correct value of $%
n_{B}(p=0)$ is obtained, then we will check the normalization condition, and
finally we will make sure that we recover the correct high-$p$ asymptotic
behavior.

\subsection{Test of $n_{B}(p)$ at zero momentum $p=0$}\label{section4a}

We wish to compare the value of momentum density $n_{B}(p)$ at $p=0$ by
using its definition in (\ref{20}) with that obtained through our general
expression (\ref{57}). While the idea of the test is simple, the
calculations are not straightforward.

Using the definition (\ref{20}) together with $\rho _{B}(x_{1},x_{2})$ given
by the first eqality in (\ref{5}), gives for $p=0$ 
\begin{equation}
n_{B}\left( 0\right) =\frac{1}{2\pi \hbar } \int_{-\infty }^{\infty }
\int_{-\infty }^{\infty } \rho _{B}\left( x_{1},x_{2}\right) dx_{1}dx_{2} = 
\frac{\ell}{\pi ^{2}\hbar } \int_{-\infty}^\infty \int_{-\infty}^\infty e^{-%
\frac{(z_1^2+z_2^2)}2} dz_1 dz_{2} \int_{-\infty}^\infty e^{-\xi^2} |z_1-\xi
| |z_2-\xi | d\xi .  \label{58}
\end{equation}%
For any fixed value of $\xi$, we make the change of variables $z_1=r+\xi$, $%
z_2=s+\xi$, and get 
\begin{equation}
n_{B}\left( 0\right) = \frac{\ell}{\pi ^{2}\hbar } \int_{-\infty}^\infty
\int_{-\infty}^\infty |r|\, |s| e^{-\frac{(r^2+s^2)}2}\, dr ds
\int_{-\infty}^\infty e^{-2\xi^2} e^{-(r+s)\xi} d\xi .  \label{59}
\end{equation}%
After the integration on $\xi $ is performed, the above expression reduces
to 
\begin{equation}
n_{B}\left( 0\right) =\frac{\ell}{\pi ^{3/2} \sqrt{2} \, \hbar }
\int_{-\infty }^\infty dr\, |r| e^{-\frac{3 }{8 }r^{2}} \int_{-\infty
}^\infty d s\, |s| e^{-\frac{3 }{8 }s^{2}} e^{\frac{r s}4} = \frac{\ell 
\sqrt{2}}{\pi ^{3/2} \hbar } \int_{-\infty }^\infty dr\, |r| e^{-\frac{3 }{8 
}r^{2}} \int_{0}^{\infty }s e^{-\frac{3}{8}s^{2}}\cosh (\tfrac{r s}4) \, ds.
\label{60}
\end{equation}
Using the identity (see eq. (3.562) number (4) of \cite{Gradshteyn--Ryzhik}) 
\begin{equation}
\int_{0}^{\infty }s e^{-\beta s^{2}}\cosh (\gamma s) ds=\frac{1}{2\beta }+%
\frac{\gamma }{4\beta }\sqrt{\frac{\pi }{\beta }}\, e^{\frac{\gamma ^{2}}{%
4\beta }}\, \text{erf}\left( \frac{\gamma }{2\sqrt{\beta }}\right)
\label{62}
\end{equation}%
for $\beta =3/8,\gamma =r/4$ in (\ref{60}), we obtain 
\begin{equation*}
n_{B}( 0) = \frac{\ell \sqrt{2}}{\pi ^{3/2} \hbar } \int_{-\infty }^\infty
|r| e^{-\frac{3 }{8 }r^{2}} dr \left[\frac{4}{3}+\sqrt{\frac{8\pi }{3}} e^{%
\frac{r^{2}}{24}} \frac{r}{6}\, \text{erf}\left( \frac{r}{4}\sqrt{\frac{2}{3}%
}\right)\right] =\frac{\ell \sqrt{2}}{\pi ^{3/2} \hbar } \left[ \frac{32}{9}+%
\sqrt{\frac{8\pi }{27}}\int_{0}^\infty r^{2}e^{-\frac{r^{2}}{3}}\, \text{erf}%
\left( \frac{r}{4}\sqrt{\frac{2}{3}}\right) \right] .
\end{equation*}%
The remaining integral can be found in the literature \cite{Ng-Gell1969},
and after simplifications we get 
\begin{equation}
n_{B}(0) =\frac{\ell}{\sqrt{\pi }\hbar }\left[ \frac{4}{\pi }\arctan ( \sqrt{%
2}) +\frac{4\sqrt{2}}{\pi }-1\right] .  \label{67}
\end{equation}%
It is interesting to compare this value of the momentum density at $p=0$
(peak momentum distribution) with its two-fermions system counterpart given
by (\ref{32}) 
\begin{equation}
n_{F}(0)=\frac{\ell}{\sqrt{\pi }\hbar } .  \label{68}
\end{equation}%
We clearly observe a large deviation measured by the ratio $n_{B}\left(
0\right) /n_{F}(0)=\left[ \frac{4}{\pi }\arctan \left( \sqrt{2}\right) +%
\frac{4\sqrt{2}}{\pi }-1\right] \allowbreak \approx 2.02$. This ratio is
featured in fig.~\ref{fig1}.

Let us now check that the result (\ref{67}) is also obtained by setting $p=0$
into expression (\ref{57}) 
\begin{equation}
n_{B} ( 0 ) =\frac{\ell}{\sqrt{\pi }\hbar }+\frac{\ell}{\sqrt{\pi }\hbar }%
\left( \frac{4\sqrt{2}}{\pi }\right) \left[ \frac{2}{3}+\Psi _{1}\left( 2,%
\frac{1}{2},\frac{3}{2},\frac{1}{2};-\frac{1}{2},0\right) -\Psi _{1}\left( 1,%
\frac{1}{2},\frac{3}{2},\frac{1}{2};-\frac{1}{2},0\right) \right] .
\label{69}
\end{equation}%
In order to compute $\Psi _{1}\left( 1,\frac{1}{2},\frac{3}{2},\frac{1}{2};-%
\frac{1}{2},0\right) $ and $\Psi _{1}\left( 2,\frac{1}{2},\frac{3}{2},\frac{1%
}{2};-\frac{1}{2},0\right) $, we return to the third form of the confluent
hypergeometric function with two variables in (\ref{52}) and set $z_{2}=0$: 
\begin{eqnarray}
\Psi _{1}\left( 1,\frac{1}{2},\frac{3}{2},\frac{1}{2};-\frac{1}{2},0\right)
={}_{2}F_{1}\left( 1,\frac{1}{2},\frac{3}{2};-\frac{1}{2}\right) , \qquad
\Psi _{1}\left( 2,\frac{1}{2},\frac{3}{2},\frac{1}{2};-\frac{1}{2},0\right)
= {}_{2}F_{1}\left( 2,\frac{1}{2},\frac{3}{2};-\frac{1}{2}\right) .
\label{71}
\end{eqnarray}%
To compute $_{2}F_{1}\left( 1,\frac{1}{2},\frac{3}{2};-\frac{1}{2}\right) $
we use the relation $_{2}F_{1}\left(1, \frac{1}{2},\frac{3}{2};-z
^{2}\right) =(\arctan z)/z$ (see (9.121) number (27) of 
ref.~\cite{Gradshteyn--Ryzhik}), with $z^2=1/\sqrt{2}$. Making use of the well known
relation $\arctan (z)+\arctan (1/z)=\pi /2$, we find 
\begin{eqnarray}
_{2}F_{1}\left( 1,\frac{1}{2},\frac{3}{2};-\frac{1}{2}\right) =\sqrt{2}%
\arctan (1/\sqrt{2}) =\sqrt{2}\left[ \frac{\pi }{2}-\arctan (\sqrt{2})\right]
.  \label{73}
\end{eqnarray}%
The other quantity $_{2}F_{1}\left( 2,\frac{1}{2},\frac{3}{2};-\frac{1}{2}%
\right)$ can be easily obtained by applying the recursion relation (15.2.14)
of \cite{Abramowitz-Stegun1972} with $a=1$, $b=1/2$ and $c=3/2$: 
\begin{eqnarray}
\text{ }_{2}F_{1}\left( 2,\frac{1}{2},\frac{3}{2};-\frac{1}{2}\right) &=&%
\frac{1}{2} \, _{2}F_{1}\left( 1,\frac{1}{2},\frac{3}{2};-\frac{1}{2}\right)
+ \frac{1}{2} \, _{2}F_{1}\left( 1,\frac{3}{2},\frac{3}{2};-\frac{1}{2}%
\right)  \notag \\
&=& \frac{1}{2} \, _{2}F_{1}\left( 1,\frac{1}{2},\frac{3}{2};-\frac{1}{2}%
\right) + \frac{1}{3}  \label{75}
\end{eqnarray}%
where the second equality is obtained by noting that $\, _{2}F_{1}\left(
1,b,b;z\right) =1/(1-z)$, here with $z=-1/2$. Substituting (\ref{71}),
together with (\ref{73}) and (\ref{75}), into (\ref{69}), we recover the
peak momentum distribution value (\ref{67}), thus ending the proof.

\subsection{Normalization condition of momentum distribution n$_{B}(p)$}

Next, we want to show that momentum density obeys the normalization
condition $\int n_{B}\left( p\right) dp=2$. It is more convenient not to use
the form of $n_{B}\left( p\right) $ in (\ref{57}), but an expression
obtained a few steps before. That is, we use the momentum density as given
in (\ref{25}), $n_{B}\left( p\right) =n_{F}(p)+ n_{D}(p)$, with the
expressions for $n_{F}(p)$ and $n_{D}(p)$ given by (\ref{30}) and 
(\ref{40}), to write 
\begin{eqnarray}
n_{B}( p) = \frac{2\ell}{\hbar \pi }\left[ \int_{-\infty }^{\infty} e^{-v^2}
\cos (2qv) dv- \int_{-\infty}^{\infty }v^{2}e^{-v^2} \cos(2qv)dv\right] +%
\frac{4\ell}{\pi ^{3/2}\hbar }\left[ \sqrt{2}\int_{-\infty}^{\infty }e^{-%
\frac{3v^{2}}{2}}|v|\cos ( 2qv) dv \right. \qquad \qquad  \notag \\
\left. -\sqrt{\pi} \int_{-\infty }^{\infty }e^{-v^2}\cos( 2q v) \, \text{erf}%
\left( \frac{|v|}{\sqrt{2}}\right) dv + \sqrt{\pi } \int_{-\infty }^{\infty}
e^{-v^2} v^{2}\cos(2qv) \, \text{erf}\left( \frac{|v|}{\sqrt{2}}\right) dv %
\right] .  \label{81}
\end{eqnarray}
Taking into account the following representation of the Dirac delta
distribution 
\begin{equation}
\int_{-\infty }^{\infty }\cos (2qv)\, dq= \pi \delta (v),  \label{83}
\end{equation}
the integral over the momentum $p$ (or equivalently over $q=\ell p/\hbar$)
is quite straightforward 
\begin{eqnarray}
\int n_{B}( p) \,dp = 2 \int_{-\infty }^{\infty } \left[ e^{-v^2} - v^{2}
e^{-v^2} + \frac{2\sqrt{2}}{\sqrt{\pi}} e^{-\frac{3v^{2}}{2}}\left \vert
v\right \vert - 2 e^{-v^2}\, \text{erf}\left( \frac{\left \vert v\right
\vert }{\sqrt{2}}\right) + 2 e^{-v^2} v^{2} \, \text{erf}\left( \frac{\left
\vert v\right \vert }{\sqrt{2}}\right) \right] \delta (v) \, dv =2,
\label{82}
\end{eqnarray}
where the final result is obtained using $\text{erf}(0)=0$. 
We have thus checked that, for the two-particle system considered in this
work, we have the correct normalization condition for $n_{B}( p) $.

\subsection{High-$p$ behavior of the momentum distribution and Tan's
coefficient}

Here, we will examine the asymptotic behavior of the momentum density $%
n_{B}(p)$ obtained in (\ref{57}) as $p\rightarrow \infty $ and we will prove
that 
\begin{equation}
n_{B}(p)\approx \frac{C_{2}}{{}p^{4}}, \quad p\rightarrow \infty , \qquad
C_{2}=(2 m\hbar \omega /\pi)^{3/2} = (2/\pi)^{3/2}\,
(\hbar/\ell)^3 ,  \label{86}
\end{equation}%
where $C_{2}$ is the exact Tan's contact coefficient for the two-atom system
under study \cite{Tan}, recovering the $1/p^{4}$ dependence originated from
the short-range character of the interaction.

To derive this result we search the asymptotic behavior up to $1/p^{4}$. For
this purpose we need the asymptotic behavior ($x\rightarrow +\infty $) \cite%
{Olver2010}%
\begin{equation}
M(a,b;-x)\approx \frac{\Gamma (b)}{\Gamma (b-a)}x^{-a}\sum
\limits_{k=0}^{\infty }\frac{(1+a-b)_{k}(a)_{k}}{k!}x^{-k} ,  \label{90}
\end{equation}%
$a\neq 0,-1,-2,...$

Taking $a=1,b=\frac{3}{2}$ and $x=2\ell^{2}p^{2}/(3\hbar ^{2})$, we have
from (\ref{90}) that
\begin{equation}
\frac{4}{3}\left( \frac{\ell p}{\hbar }\right) ^{2}M\left( 1,\frac{3}{2};-%
\frac{2}{3}\frac{\ell^{2}p^{2}}{\hbar ^{2}}\right) \approx \sum
\limits_{k=0}^{\infty }\frac{(\frac{1}{2})_{k}(1)_{k}}{k!}\left( \frac{%
3\hbar ^{2}}{2\ell^{2}p^{2}}\right) ^{k} =1+\frac{3}{4} \frac{\hbar ^{2}}{%
\ell^{2}p^{2}}+\frac{27}{16} \frac{\hbar ^{4}}{\ell^{4}p^{4}}+O\left( \frac{1%
}{p^{6}}\right) .  \label{93}
\end{equation}

To determine the asymptotic behavior of $\Psi _{1}\left(
1+j,1/2,3/2,1/2;-1/2,-\ell^{2}p^{2}/\hbar ^{2}\right)$ as $p\rightarrow
\infty $, with either $j=0$ or $j=1$, we use the second expression in %
(\ref{50}) to write 
\begin{equation}
\Psi _{1}\left( 1+j,\frac{1}{2},\frac{3}{2},\frac{1}{2};-\frac{1}{2},-\frac{%
\  \ell^{2}p^{2}}{\hbar ^{2}}\right) =\sum \limits_{m=0}^{\infty }\frac{%
(1+j)_{m}(\frac{1}{2})_{m}}{(\frac{3}{2})_{m}}\frac{\left( -\frac{1}{2}%
\right) ^{m}}{m!}M\left( 1+j+m,\frac{1}{2};-\frac{\  \ell^{2}p^{2}}{\hbar ^{2}%
}\right) \text{ }  \label{95}
\end{equation}%
and then the asymptotic expansion (\ref{90}) to get 
\begin{eqnarray}
\Psi _{1}\left( 1+j,\frac{1}{2},\frac{3}{2},\frac{1}{2};-\frac{1}{2},-\frac{%
\ell^{2}p^{2}}{\hbar ^{2}}\right) \approx \sum \limits_{m=0}^{\infty }&&%
\frac{(1+j)_{m}(\frac{1}{2})_{m}}{(\frac{3}{2})_{m}}\frac{\left( -\frac{1}{2}%
\right) ^{m}}{m!}\frac{\Gamma (\frac{1}{2})}{\Gamma (-m-j-\frac{1}{2})}\
\left( \frac{\ell^{2}p^{2}}{\hbar ^{2}}\right) ^{-m-j-1}  \notag \\
&&\times \sum \limits_{k=0}^{\infty }\frac{(m+j+\frac{3}{2})_{k}(m+j+1)_{k}}{%
k!}\left( \frac{\ell^{2}p^{2}}{\hbar ^{2}}\right) ^{-k}.  \label{96}
\end{eqnarray}%
To retain only terms up to $1/p^{4}$, we have to consider the summation
indexes up to $m+k\leq 1-j$. It is easy to show that 
\begin{eqnarray}
\Psi _{1}\left( 1,\frac{1}{2},\frac{3}{2},\frac{1}{2};-\frac{1}{2},-\frac{%
\ell^{2}p^{2}}{\hbar ^{2}}\right) &\approx& -\frac{\hbar ^{2}}{2\ell^{2}p^{2}%
}-\frac{7}{8}\frac{\hbar^{4}}{\ell^{4}p^4}+O\left( \frac{1}{p^{6}}\right).
\label{98} \\
\Psi _{1}\left( 2,\frac{1}{2},\frac{3}{2},\frac{1}{2};-\frac{1}{2},-\frac{%
\ell^{2}p^{2}}{\hbar ^{2}}\right) &\approx& \frac{3}{4}\frac{\hbar ^{4}}{%
\ell^{4}p^{4}}+O\left( \frac{1}{p^{6}}\right) .  \label{100}
\end{eqnarray}

Collecting expressions (\ref{93}), (\ref{98}) and (\ref{100}) into (\ref{57}), 
after the cancellation of the terms of order $1/p^{2}$, one finds 
\begin{equation}
n_{B}\left( p\right) \approx \left( \frac{2\hbar ^{3}\sqrt{2}}{\ell^{3}\pi ^{%
\frac{3}{2}}}\right) \frac{1}{p^{4}}= \left( \frac{2m\hbar \omega }{\pi}%
\right)^{\frac{3}{2}} \frac{1}{p^{4}},  \label{102}
\end{equation}
which is the result announced in (\ref{86}): we have recovered the well
known $1/p^4$ asymptotic decay of $n_{B}(p)$ for high momentum values. The
coefficient in front of the \textit{high}-$p$ tail of the momentum
distribution, called Tan's coefficient \cite{Tan}, is given by $(2/\pi
)^{3/2}(m\omega \hbar )^{3/2}$ for the system of two particles under study.
Note, finally, that from a basic dimensional analysis, expressions in %
(\ref{102}) or (\ref{86}) have correct units of momentum density.

\begin{figure}[htb]
\centering
\includegraphics[width=0.45\textwidth]{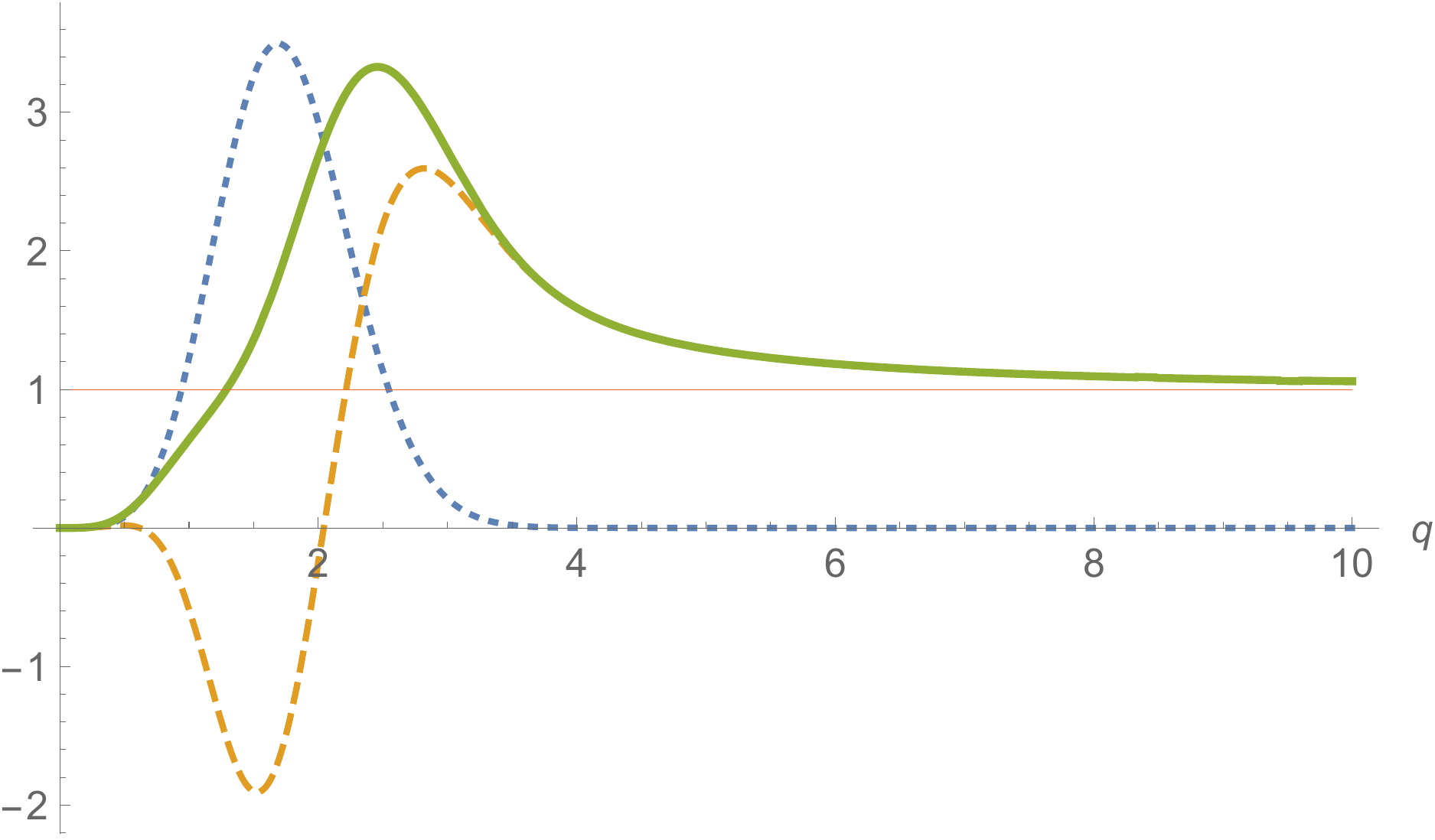}
\caption{Plot of $n_{B}(p) p^4/C_2 $ (green solid curve), $%
n_{F}(p) p^4/C_2 $ (blue dotted curve) and $n_{D}(p) p^4/C_2 $ (yellow
dashed curve) as a function of $q=\ell p/\hbar$, with $C_2= (2/\protect \pi%
)^{3/2}\, (\hbar/\ell)^3$.}
\label{fig2}
\end{figure}

The high-$p$ tail behavior can be observed in fig.~\ref{fig2} where we
show $n_{B}(p), n_{F}(p)$ and $n_{D}(p)$ multiplied by $p^4/C_2$. One can
appreciate how $n_{B}(p)$ tends to 1 for large $q$, and that the asymptotic
limit is dominated by the deviation term $n_D$, since the $n_F$ is already
negligible for $q>4$. 

\section{Conclusion}

In this paper we have studied a one-dimensional quantum system consisting of
two bosons harmonically trapped and with an infinite contact repulsion. This
system has been the subject of numerous studies, in particular with respect
to its asymptotic momentum distribution. In the present work we have been
able to provide an exact expression of the momentum distribution valid for
arbitrary value of the momentum, and thus we have extended the list of
exactly solvable models of two interacting particles. Starting from the wave
function of the two particles, we have derived an exact expression of the
one-body density matrix expressed in terms of centre of mass and relative
coordinates. From this result we have calculated the corresponding momentum
distribution. To clearly identify the deviation of this system from the two
noninteracting fermions, we wrote this momentum density as the sum of two terms:
a first one corresponding to the two noninteracting fermions, and a second
one related to the short range character of the interaction.

In order to validate our analytic expression of the momentum density, three
robust tests have been performed satisfactorily. In particular, the third
test concerns the asymptotic decay for high momentum values; we have been
able to derive the exact expression of the Tan contact coefficient, which we
recall is the coefficient in front of the high-$p$ tail of the momentum
distribution, for the case of two particles \cite{Tan}. 

\acknowledgements
This research was funded by Junta de Castilla y Le\'on and FEDER projects VA137G18 and BU229P18.

\appendix

\section{}
\label{appendixa}

In this Appendix we will calculate the two functions $\mathcal{B}(u,v)$ and $%
\mathcal{C}(u,v)$ given by (\ref{11}), and thereafter $\mathcal{A}(u,v)$
defined in (\ref{7}). Making the change of variable $t=w+u$ in $\mathcal{B}%
(u,v)$, we get 
\begin{equation}
\mathcal{B}(u,v)= \int_{-\infty }^{u-\left \vert v\right \vert} t^{2}
e^{-t^2}\, dt -2u\int_{-\infty }^{u-\left \vert v\right \vert}\,
te^{-t^{2}}\, dt +\left( u^{2}-v^2\right) \int_{-\infty}^{u-\left \vert
v\right \vert} e^{-t^{2}}\, dt .  \label{A1}
\end{equation}
If we perform an integration by parts in the first integral, we then get
after rearrangements 
\begin{equation}
\mathcal{B}(u,v)=\frac12 \left( u+|v| \right) e^{-(u-|v|)^{2}}+\left(
\tfrac12 +u^{2}-v^2\right) \int_{-\infty }^{u-|v|}\, e^{-t^{2}}\, dt .
\label{A2}
\end{equation}
Proceeding similarly for $\mathcal{C}(u,v)$, we write 
\begin{equation}
\mathcal{C} (u,v)=\int_{u-\left \vert v\right \vert }^{u+\left \vert v\right
\vert}\, t^{2}e^{-t^{2}}\, dt-2u \int_{u-\left \vert v\right \vert}^{u+\left
\vert v\right \vert}\, te^{-t^{2}}\, dt+\left( u^2-v^2\right) \int_{u-\left
\vert v\right \vert}^{u+\left \vert v\right \vert}\, e^{-t^{2}}\, dt .
\label{A5}
\end{equation}%
Carrying out an integration by parts of the first integral, transforms %
(\ref{A5}) as 
\begin{eqnarray}
\mathcal{C} (u,v) &=&\frac12 \left( u-\left \vert v\right \vert \right)
e^{-(u+\left \vert v\right \vert )^{2}}-\frac12\left( u+{\left \vert v\right
\vert } \right) e^{-(u-{\left \vert v\right \vert })^{2}} + \left(
\frac12+u^{2}-v^2\right) \int_{u-\left \vert v\right \vert}^{u+\left \vert
v\right \vert} e^{-t^{2}}\, dt .  \label{A6}
\end{eqnarray}%
Collecting the two results (\ref{A2}) and (\ref{A6}), the expression of $%
\mathcal{A} (u,v)$ in (\ref{7}) becomes 
\begin{eqnarray}
\mathcal{A} (u,v) = (u+{|v|})e^{-(u-{|v|})^{2}}-(u-{|v|})e^{-(u+{|v|})^{2}}
+ ( \tfrac12+u^{2}-v^2) \left[ \int_{-\infty }^{u-\left \vert v\right \vert}
e^{-t^{2}} dt+\int_{-\infty }^{-u-\left \vert v\right \vert}e^{-t^{2}}
dt-\int_{u-\left \vert v\right \vert}^{u+\left \vert v\right
\vert}e^{-t^{2}} dt\right] .  \notag
\end{eqnarray}%
To further simplify, let us observe that 
\begin{eqnarray}
\int_{-\infty }^{u-\left \vert v\right \vert} e^{-t^{2}} dt+\int_{-\infty
}^{-u-\left \vert v\right \vert}e^{-t^{2}} dt =\int_{-\infty }^{\infty
}e^{-t^{2}}dt-\int_{u-\left \vert v\right \vert}^{u+\left \vert v\right
\vert}e^{-t^{2}}dt = \sqrt{\pi}-\int_{u-\left \vert v\right \vert}^{u+\left
\vert v\right \vert}e^{-t^{2}}\, dt .  \label{A9}
\end{eqnarray}%
Using the definition of the $\text{erf}$ function $\int%
\nolimits_{0}^{a}e^{-t^{2}}dt=\frac{1}{2}\sqrt{\pi}\, \text{erf}(a)$, we
further have 
\begin{equation}
\int_{u-\left \vert v\right \vert}^{u+\left \vert v\right \vert}e^{-t^{2}}
dt =\frac{\sqrt{\pi}}{2} \bigl( \text{erf}\left( u+\left \vert v\right \vert
\right) -\text{erf} \left( u-\left \vert v\right \vert \right) \bigr).
\label{A10}
\end{equation}
Combining the above two results, the expression of $\mathcal{A} (u,v)$
becomes 
\begin{equation}
\mathcal{A} (u,v) = (u+{|v|})e^{-(u-{|v|})^{2}}-(u-{|v|})e^{-(u+{|v|})^{2}} +%
\sqrt{\pi} ( \tfrac12+u^{2}-v^2) \bigl( 1+\text{erf}\left( u-\left \vert
v\right \vert \right) -\text{erf} \left( u+\left \vert v\right \vert \right) %
\bigr) .  \label{A11}
\end{equation}

\section{}
\label{appendixb}

In this Appendix we consider the double integral $I_2$ in %
(\ref{38})
\begin{equation}
I_2= \frac{4\ell}{\pi ^{3/2}\hbar }\int_{-\infty }^{\infty
}e^{-v^2} \cos(2qv)\, dv \int_{-\infty }^{\infty }\left( \tfrac{1}{2}-v^2
+u^{2}\right) \, \text{erf}\left( u-|v|\right) \, e^{-u^{2}}du = \frac{4\ell%
}{\pi ^{3/2}\hbar } \, ( K_{1}+K_{2}+K_{3}),  \label{B1}
\end{equation}
which, for convenience, we have split into three double integrals defined
through 
\begin{eqnarray}
K_{1} &=&\frac{1}{2}\int_{-\infty }^{\infty }e^{-{v^{2}}}\cos(2qv)dv\int_{-%
\infty }^{\infty }e^{-u^{2}}\text{erf}\left( u-{|v|}\right) du ,  \label{B2}
\\
K_{2} &=&- \int_{-\infty }^{\infty }v^{2}e^{-{v^{2}}}\cos(2qv)dv\int_{-%
\infty }^{\infty }e^{-u^{2}}\text{erf}\left( u-{|v|} \right) du ,  \label{B3}
\\
K_{3} &=&\int_{-\infty }^{\infty }e^{-{v^{2}}}\cos(2qv)dv\int_{-\infty
}^{\infty }u^{2}e^{-u^{2}}\text{erf}\left( u-{|v|}\right) du .  \label{B4}
\end{eqnarray}
In order to compute the $u$ integral in (\ref{B4}), we first perform an
integration by parts to get 
\begin{equation}
G=\int_{-\infty }^{\infty }u^{2}e^{-u^{2}}\text{erf}\left( u-|v|\right) du=%
\frac{1}{2}\int_{-\infty }^{\infty }e^{-u^{2}}\text{erf}\left( u-|v|\right)
du+\frac{1}{\sqrt{\pi }}\int_{-\infty }^{\infty }e^{-u^{2}}ue^{-\left( u-
|v|\right) ^{2}}du .  \label{B9}
\end{equation}
The second integral on the right-hand side of (\ref{B9}) can be carried out,
giving 
\begin{equation}
G=\frac{1}{2}\int_{-\infty }^{\infty }e^{-u^{2}}\text{erf}\left(
u-|v|\right) du+ \frac{|v|}{2\sqrt{2}}e^{-|v|^2/2}.  \label{B11}
\end{equation}
Hence, substituting (\ref{B11}) into (\ref{B4}), one finds out that $K_3$ is
related to $K_1$ through 
\begin{equation}
K_{3}=K_{1}+\frac{1}{\sqrt{2}}\int_{0}^\infty e^{-3v^2/2} \,v\, \cos(2qv)\,
dv .  \label{B12}
\end{equation}

To carry out the $u$ integrals in (\ref{B2})--(\ref{B3}), we start from the
identity 
\begin{equation}
\int_{-\infty }^{\infty }e^{-(x+a)^{2}}\text{erf}\left( x\right) dx=-\sqrt{%
\pi}\, \text{erf}\left( \frac{a}{\sqrt{2}}\right) =\int_{-\infty }^{\infty
}e^{-t^{2}}\text{erf}\left( t-a\right) dt.  \label{B5}
\end{equation}
Using the above relation for $a=|v|$ allows us to write the expressions of %
(\ref{B2}) and (\ref{B3}) as single integrals, so that 
\begin{eqnarray}
K_{1} =-\sqrt{\pi }\int_{0}^{\infty }e^{-{v^{2}}}\cos (2qv)\, \text{erf}%
\left( \frac{v}{\sqrt{2}}\right) dv, \qquad K_{2} = 2\sqrt{\pi}
\int_{0}^\infty v^{2}e^{-v^2}\cos (2qv)\, \text{erf}\left( \frac{v}{\sqrt{2}}%
\right) dv.  \label{B8}
\end{eqnarray}

Finally adding the three results (\ref{B8}) and (\ref{B12}), we can write (\ref{B1}) as
\begin{eqnarray}
I_2
&=& \frac{4\ell}{\pi ^{3/2}\hbar }\left \{ -2\sqrt{\pi }\int_{0}^{\infty
}e^{-v^{2}}\cos (2qv)\, \text{erf}\left( \frac{v}{\sqrt{2}}\right) dv \right.
\label{B14} \\
&&\qquad\qquad \left. +2\sqrt{\pi} \int_{0}^{\infty }v^{2}e^{-v^2}\cos (2qv)\, 
\text{erf}\left( \frac{v}{\sqrt{2}}\right) dv + \frac{\sqrt{2}}{2}%
\int_{0}^{\infty }e^{-3v^2/2}\,v\, \cos (2qv)\, dv \right \}.  \notag
\end{eqnarray}

\end{document}